\documentclass[twocolumn,prd,showpacs,preprintnumbers,amsmath,amssymb]{revtex4}

\usepackage{graphicx}
\usepackage{dcolumn}
\newcommand{\figref}[2][{}]{\hyperref[#2]{\figurename~\ref{#2}#1}}
\usepackage{bm}
\usepackage{epsfig}
\usepackage{epstopdf}
\usepackage{grffile}
\usepackage{color}
\usepackage{colordvi}
\usepackage{amsmath,amssymb}
\usepackage{rotating}
\usepackage{lscape}
\usepackage{float}
\usepackage{multirow}
\usepackage{footnote}
 \makesavenoteenv{environmentname}

\usepackage{hyperref}

\usepackage[T1]{fontenc}

\begin{document}

\title{ Intrinsic strange distributions in the nucleon from the light-cone models}

\author{Maral Salajegheh}

\email{Maral.Salajegheh@gmail.com}

\affiliation{Department of Physics, Kerman Branch, Islamic Azad University, Kerman, Iran}

\date{\today}

\begin{abstract}
Precise knowledge of the strange and antistrange quark distributions of the
 nucleon is a major step toward better understanding of the strong interaction
 and the nucleon structure. Moreover, the $ s-\bar s $ asymmetry in the nucleon 
 plays an important role in some  physical processes involving hadrons. 
 The goal of this paper is the study of intrinsic strange contribution to the 
 strange sea of the nucleon. To this aim, we calculate the intrinsic 
 strange distributions from the various light-cone models, including Brodsky, Hoyer, Peterson, and Sakai (BHPS); 
 scalar five-quark and meson-baryon models and then compare their results. 
 These models can lead to the rather different distributions for the intrinsic 
 strange that are dominated in different values of $ x $. Furthermore, 
 the meson-baryon model leads to the $ s-\bar s $ asymmetry that can be 
 comparable in some situations to the result obtained from the global analysis of PDFs. 
 We also present a simple parametrization for each model prediction of intrinsic strange distribution.
\end{abstract}

\maketitle

\section{INTRODUCTION}\label{int} 
For more than three decades, the intrinsic quark content of the nucleon has 
been a subject of interest in many studies in hadron physics. 
The existence of a nonperturbative quark component in the nucleon 
that is natural in the light-cone Fock space picture was suggested by 
Brodsky, Hoyer, Peterson, and Sakai (BHPS) \cite{Brodsky:1980pb, 1981} 
for the first time. According to the BHPS model, there are two 
distinct types of quark contributions to the nucleon sea: ``extrinsic'' sea 
quarks that are produced perturbatively in the splitting of gluons into $ q\bar q $ 
pairs in the DGLAP $ Q^2 $ evolution \cite{Gribov:1972ri} and ``intrinsic'' sea 
quarks that are originated through the nonperturbative fluctuations of the nucleon 
state to five-quark states or a meson plus baryon state. One of the main differences 
between extrinsic and intrinsic sea quarks is that they have different treatments 
(``sealike'' and valencelike'' respectively) and then dominate in different regions 
of momentum fraction $ x $.

One can perhaps divide all of the studies that have been performed so far in the 
intrinsic quark subject into two main categories: heavy and light intrinsic quarks. 
Both categories contain theoretical calculations from the various light-cone models, 
extraction of the probabilities of the intrinsic quarks in the nucleon using available 
experimental data, and impact of intrinsic quarks on the physical observables sensitive 
to their distributions. For example, in addition to the BHPS result for the intrinsic 
charm (IC) distributions in the nucleon, there have been some other theoretical calculations 
using the scalar five-quark and meson-baryon models performed by 
Pumplin \cite{Pumplin:2005yf} and Hobbs \textit{et al.} \cite{Hobbs:2013bia} in recent years. 
In the case of intrinsic light quark distributions, some calculations have been 
performed by Chang and Pang \cite{Chang:2011vx} using the BHPS model and 
there are also a wide range of results from the meson cloud model (MCM) 
\cite{Thomas:1983fh, Signal:1987gz, Melnitchouk:1992yd, Brodsky:1996hc, 
Szczurek:1996ur, Speth:1996pz, Christiansen:1998dz, Kumano:1997cy, 
Melnitchouk:1998rv, Cao:1999da, Cao:2003ny, Cao:2003zm, 
Chen:2009xy, Traini:2013zqa} and chiral quark model (CQM) \cite{Manohar:1983md, 
Eichten:1991mt, Szczurek:1996tp, Song:1997bp, Wakamatsu:1998rx, Cheng:1997tt, 
Wakamatsu:2003wg, Wakamatsu:2014asa} for either polarized or 
unpolarized distributions. For the extraction of the probabilities of the intrinsic 
charm quark in the nucleon, we can point to the analyses performed by BHPS 
\cite{1981} using diffractive production of charmed hadrons; by Harris \textit{et al.} 
\cite{Harris:1995jx} using the EMC charm production data \cite{Aubert:1982tt}; 
and also some global analyses of parton distribution functions (PDFs) performed 
so far, including the intrinsic charm contributions in the nucleon \cite{Pumplin:2007wg,
Nadolsky:2008zw, Jimenez-Delgado:2014zga}. Moreover, some analyses have been 
done to extract the probabilities of the light intrinsic quarks \cite{Chang:2011vx, 
Chang:2011du} using the existing $ \bar d - \bar u $ data from the Fermilab E866 
Drell-Yan experiment \cite{Towell:2001nh} and $ s+\bar s $ data from the HERMES 
Collaboration measurement of charged kaon production in the 
semi-inclusive deep inelastic scattering (SIDIS) reaction 
\cite{Airapetian:2008qf}. There have also been some studies about the impact of the intrinsic charm 
quark on processes such as direct photon \cite{Bednyakov:2014pqa} and $ Z $ boson 
\cite{Beauchemin:2014rya} production in association with a heavy quark or intrinsic bottom 
quark on heavy new physics \cite{Lyonnet:2015dca} at the present hadron colliders such as LHC.

Although the heavy quarks have an important role in the study of many 
processes in the standard model and beyond it, one of the 
significant aspects of the nucleon structure is the distribution of strange 
and antistrange sea quarks and their possible asymmetry. More precise 
knowledge in this field is very important for better understanding of the 
nucleon structure and properties of the sea quarks and also for describing processes such as $ W $ boson production in association with 
charm jets \cite{Abazov:2014fka} or a single top quark production 
\cite{He:2011ss}, as well as neutrino interactions \cite{Alberico:2001sd, 
Dore:2011qe}. In the present study, we concentrate on the intrinsic 
strange quark and calculate its distribution in the nucleon using 
various light-cone models. Actually, in addition to using the BHPS model 
as has been done in Ref. \cite{Chang:2011vx}, we use the scalar five-quark 
model and a simple meson-baryon model (MBM) introduced by Pumplin 
\cite{Pumplin:2005yf} (and applied for the intrinsic charm) to calculate the 
intrinsic strange distribution in the nucleon numerically and then compare the 
obtained results with each other. These models can lead to the rather different 
distributions for the intrinsic strange that can be dominated in different values 
of $ x $. Although the BHPS and scalar five-quark models cannot give us any 
asymmetry between the $ s $ and $ \bar s $ distributions in the nucleon, the MBM leads to the $ s-\bar s $ asymmetry. This is a very important 
conclusion because the perturbative (extrinsic) sea quark distributions in the 
nucleon are clearly symmetric, so the flavor asymmetry observed in the nucleon 
strange sea \cite{Mason:2007zz} certainly has a nonperturbative origin. On the 
other hand, this asymmetry can also be very important for explaining some 
experimental results such as the NuTeV anomaly reported by the NuTeV 
Collaboration \cite{Zeller:2001hh}. It should be noted that the $ s-\bar s $ 
asymmetry can result also from the chiral quark model \cite{Wakamatsu:2014asa}.

The content of the present paper goes as follows: We describe briefly 
the light-cone picture of the nucleon and review the BHPS and scalar 
five-quark models in Sec.~\ref{five}. Then we present and compare 
the obtained numerical results for the intrinsic strange distribution 
using these models at the end of this section. A simple meson-baryon 
model used by Pumplin to calculate the intrinsic charm distributions 
\cite{Pumplin:2005yf} is introduced in Sec.~\ref{MBM} and is used to 
calculate the intrinsic strange distributions. The calculations are done for 
two states $ K\Lambda $ and $ K^*\Lambda $ and the resulting $ s-\bar s $ 
asymmetry from each of these states, and their sums are also presented. 
In Sec.~\ref{result}, we discuss the probability of the intrinsic strange 
in the nucleon and compare the obtained result for the $ s-\bar s $ asymmetry 
from the meson-baryon model presented in Sec.~\ref{MBM} to the NNPDF 
\cite{Ball:2012cx} result for this quantity. We summarize our results and 
conclusions in Sec.~\ref{summery}. A simple parametrization for each 
model prediction calculated in this work is given in Appendix. 
\section{Five-Quark Models in the light-cone frame}\label{five}
The light-cone frame is very useful to understand the internal structure 
of hadrons that is one of the most interesting subjects of nuclear and 
particle physics \cite{Brodsky:2004tq}. Actually, since in the light-cone 
Fock space picture the physical vacuum state has a much simpler structure, 
the light-cone wave functions provide a perfect description of hadrons that is 
frame and process independent.

In other words, if we define the proton state at fixed light-front time, 
we find that it is natural to expect nonperturbative intrinsic 
quark and gluon components in the proton wave function. 
To be more precise, we cannot consider the proton as a three-quark bound state  $\vert uud \rangle$ and its wave function is 
a superposition of quark and gluon Fock states such as $\vert uudg 
\rangle$, $\vert uudq\bar{q} \rangle$, etc., or in a more 
dynamical way provided by the meson-baryon models, a superposition of 
configurations of off-shell physical particles.
As can be seen, one of these states is the five-quark state 
$\vert uudq\bar{q} \rangle$. Although one can find a review 
in Refs.~\cite{Pumplin:2005yf, Hobbs:2013bia}, in the next 
two subsections we present briefly two models that can give 
us $q$ and $\bar{q}$ distributions in the five-quark state 
$\vert uudq\bar{q} \rangle$. However, these models cannot give us any information about the magnitude of the 
probability of this state in the proton and it should be estimated in other ways. We introduce the meson-baryon models in Sec.~\ref{MBM} separately.
\subsection{The BHPS model}\label{brodsky}
The simplest five-quark model is the BHPS model which was 
proposed in 1980 by Brodsky \textit{et al.} \cite{Brodsky:1980pb}. Actually, they found that considering 
a significant $\vert uudc\bar{c} \rangle$ Fock component in 
the proton can lead to enhanced production of charmed hadrons 
and explain their unexpected large production rates at the forward 
rapidity region. According to the BHPS model, for a $\vert uud Q\bar{Q}\rangle $  
Fock state of the proton where $Q$ is a heavy quark, neglecting the effect of the
transverse momentum in the five-quark transition amplitudes, the momentum 
distributions of  the constituent quarks are given by \cite{1981}
\begin{equation}\label{1}
P(x_1,...,x_5) = N \frac{\delta \left(1-\sum\limits_{i=1}^5 x_i\right)}{\left(m_p^2- \sum\limits_{i=1}^5 \frac{m_i^2}{x_i}\right)^2},
\end{equation}
where $m_p$ is the mass of the proton and $m_i$ and $x_i$ are 
the mass of quark $i$ and momentum fraction carried by it in the 
five-quark Fock state.  The momentum conservation is satisfied by 
virtue of the delta function. $N$ is the normalization factor and is
determined from $ \int dx_1...\times dx_5 P(x_1,...,x_5)\equiv \mathcal{P}^{Q\bar{Q}}_5$, 
where $\mathcal{P}^{Q\bar{Q}}_5$ 
is the probability of the $\vert uud Q\bar{Q}\rangle $ Fock state in 
the proton and expected to be roughly proportional to $1/m_Q^2$. 
Integrating Eq.~\eqref{1} over $x_1, x_2, x_3,$ and $x_4$, 
we get the $\bar{Q}$ distribution in the proton.

In the case of intrinsic charm, BHPS considered another 
simplifying assumption (in addition to neglecting the 
effect of the transverse momentum) that the charm
mass is much greater than the nucleon and light quark
masses [$m_4^2=m_5^2 \gg m_p^2,m_i^2(i=1,2,3)$]. 
In this way, the momentum distribution of $\vert uud c\bar{c}\rangle $ 
Fock state becomes
\begin{equation}\label{2}
P(x_1,...,x_5) = N_5 \frac{x_4^2 x_5^2}{(x_4 + x_5)^2}\delta \left(1-\sum_{i=1}^5 x_i\right),
\end{equation}
where $N_5=N/m_{4,5}^4$. Now, this equation can be solved analytically 
so that by carrying out all of the integrals except one (over $x_5$) the $x$ distribution 
of the intrinsic charm in the proton is obtained as follows
\begin{align}\label{3}
P(x_5) &= \frac{1}{2}N_5x_5^2 \Big[\frac{1}{3}(1 - x_5) (1+ 10 x_5 + x_5^2)\nonumber\\
&\quad-2 x_5 (1 - x_5)\ln (1/x_5)\Big].
 \end{align}
By performing one more integration over $x_5$ we 
obtain $\mathcal{P}^{c\bar{c}}_5=N_5/3600$. So far, 
several studies have been done to estimate the 
magnitude of $\mathcal{P}^{c\bar{c}}_5$ (or momentum 
fraction carried by intrinsic charm) and have suggested different 
values for it \cite{1981, Harris:1995jx, Nadolsky:2008zw, 
Jimenez-Delgado:2014zga}. For a review on the theoretical 
calculations, constraints from global analyses, and
collider observables sensitive to the intrinsic heavy quark 
distributions, see Ref. \cite{Brodsky:2015fna}.

As mentioned above, for the case of intrinsic heavy quarks, 
$\mathcal{P}^{Q\bar{Q}}_5$ is proportional to $1/m_Q^2$. 
Although this dependence is not applicable for the intrinsic light 
quarks, we expect that the light five-quark states $\vert uud u
\bar{u}\rangle $, $\vert uud d\bar{d}\rangle $, and $\vert uud s
\bar{s}\rangle $ have larger probabilities than the $\vert uud c\bar{c}\rangle $ 
state in the proton. In recent years, Chang and Pang, generalized 
the BHPS model to the light five-quark states  \cite{Chang:2011vx, 
Chang:2011du}. Actually, in addition to calculating the intrinsic light 
quark distributions in the proton, they extracted the probabilities 
of these states ($\mathcal{P}^{q\bar{q}}_5$) using available 
experimental data. Since the main purpose of this work is to 
calculate the intrinsic strange distribution in the proton from the various 
light-cone models and to compare them with one another, we also calculate 
the $x$ distribution of the intrinsic strange quark in the $\vert uud s\bar{s}
\rangle $ Fock state from the BHPS model. We present the obtained 
results from the five-quark models at the end of this section.
 \subsection{Scalar five-quark model}\label{pump}
 Another light-cone model that can be used to extract the $x$ distribution 
 of $Q$ in the $\vert uud Q\bar{Q}\rangle $ Fock state is the scalar 
 five-quark model. In this model that was presented by Pumplin 
 \cite{Pumplin:2005yf} during the study of intrinsic heavy quark 
 probability in the proton, the light-cone probability distributions 
 derive directly from Feynman diagram rules and some simplifying 
 assumptions that were considered in the BHPS model are eliminated. 
 According to the scalar five-quark model, if a point scalar particle with mass
$ m_0 $ and spin $0$ couples to $ N $ scalar particles with masses 
$ m_1, m_2, ..., m_N $ and spin $ 0 $ by a point-coupling $ ig $, then the 
probability distribution $ dP $ can be written as
\begin{align}\label{4}
dP&= \frac{g^2}{(16\pi^2)^{N-1}(N-2)!}\,
    \prod_{j=1}^N dx_j\, \delta\left( 1-\sum_{j=1}^N x_j\right)\nonumber\\
  &\quad\times\int_{s_0}^\infty ds\, \frac{(s-s_0)^{N-2}}{(s-m_0^2)^2}\,
    |F(s)|^2,
\end{align}
where
\begin{equation}\label{5}
s_0 = \sum_{j=1}^N \frac{m_j^2}{x_j}\,.
\end{equation}
Although high mass Fock states are regulated by the factor 
$(s-m_0^2)^{-2}$ in this model, in order to include the 
effects of the finite size of the proton we should consider further 
suppression of high-mass states to make the model more realistic 
and the integrated probability converge. In this regard, we can 
include the form factor $F^2$ as a function of $ s $ that characterizes 
the dynamics of the bound state to suppress the contributions from the
high-mass configurations. Pumplin suggested two exponential and power-law
forms for wave function factor $ F^2 $ as follows:
\begin{equation}\label{7}
|F^2(s)| =\exp [-(s-m_0^2)/\Lambda^2],
\end{equation}
\begin{equation}\label{8}
|F^2(s)|= (s +\Lambda^2 ) ^ {-n},
\end{equation}
where $\Lambda$ is a cutoff mass regulator and any 
value between 2 and 10 GeV can be chosen for it.
   \begin{figure}[t!]
    \includegraphics[width=0.45\textwidth,clip]{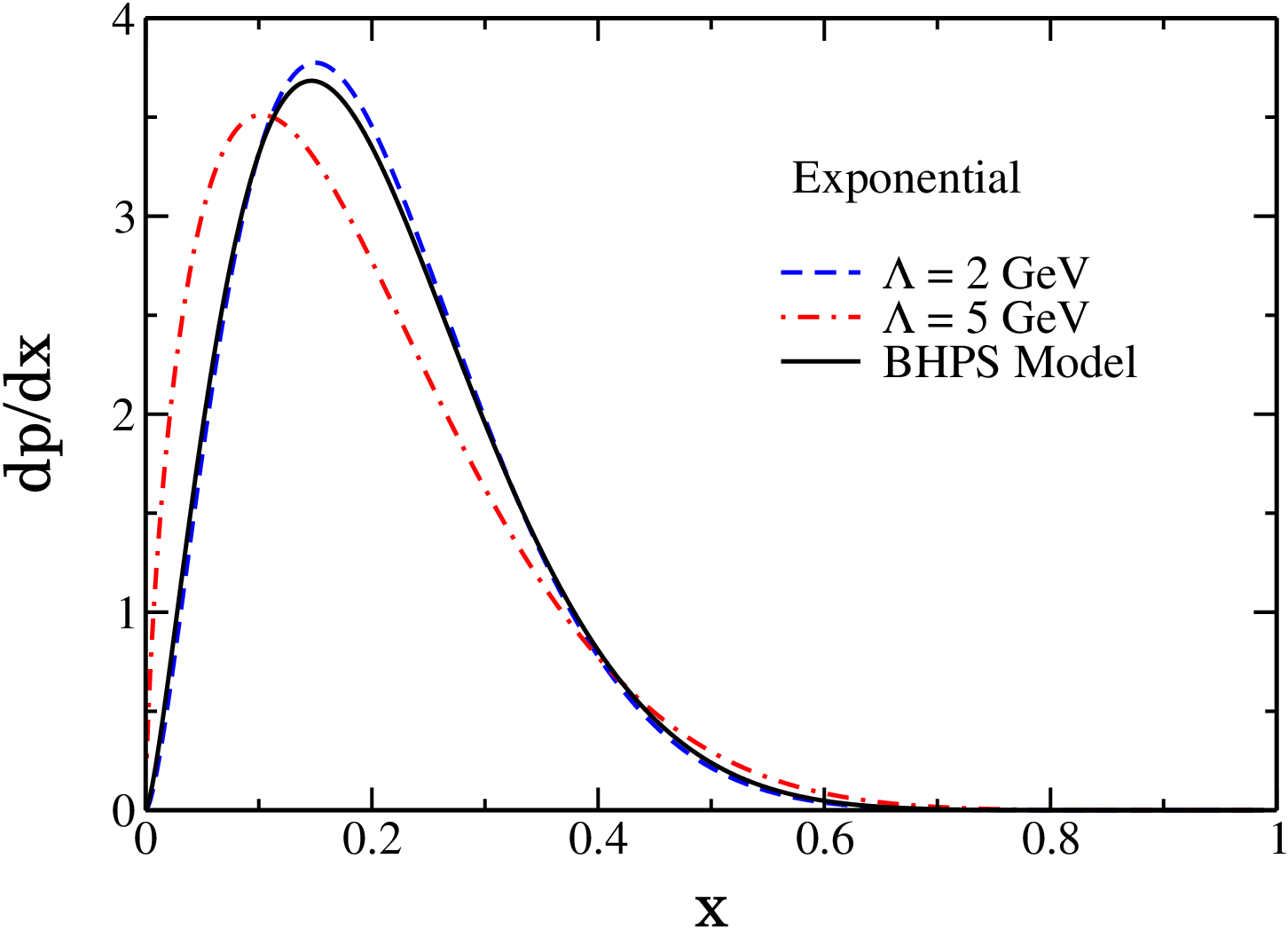}
    \includegraphics[width=0.45\textwidth,clip]{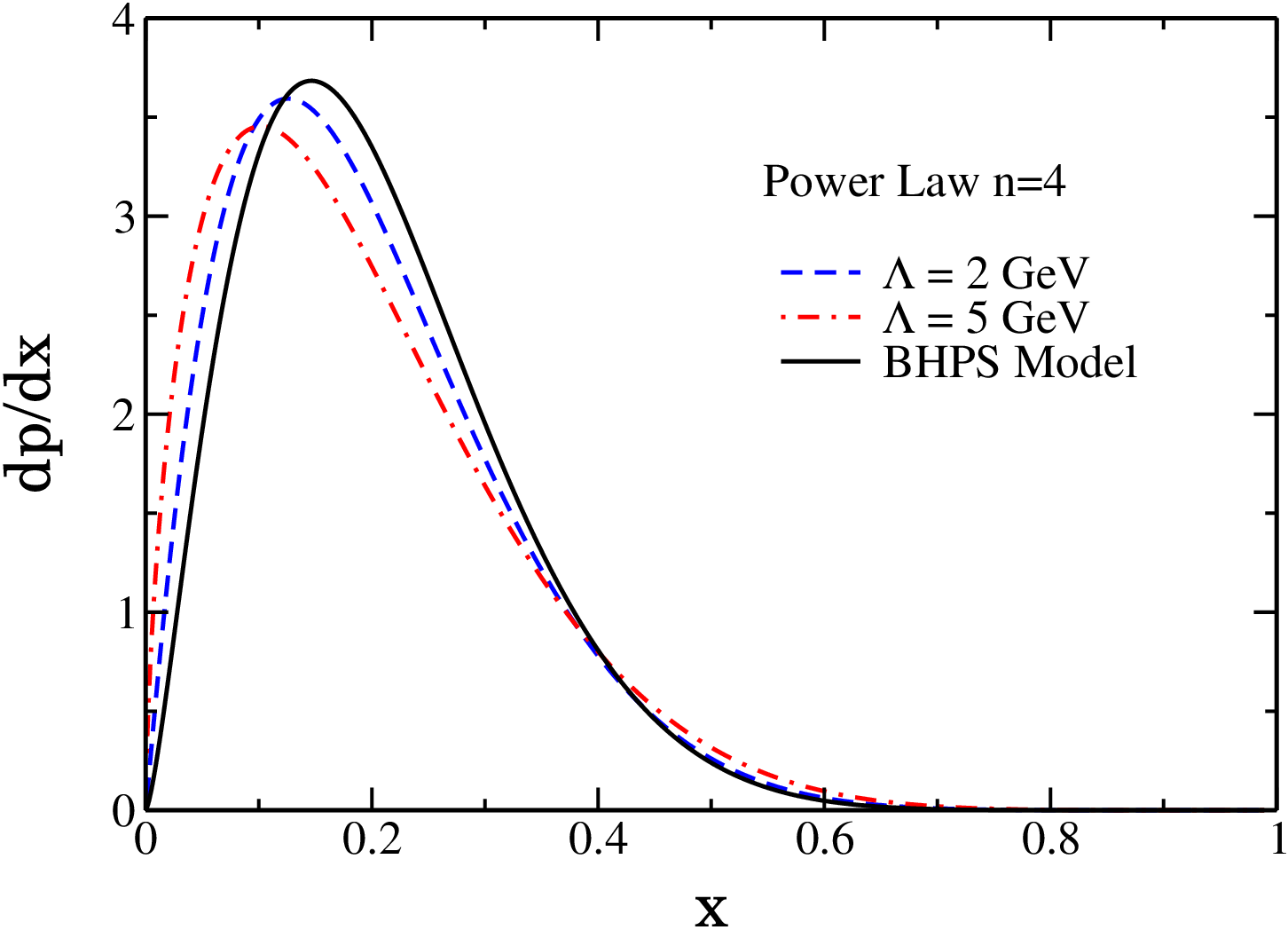}
    \caption{ The intrinsic strange distribution from the BHPS 
    and scalar five-quark model with the exponential (top) and power-law (bottom) form factor. }
    \label{pumpli}
\end{figure}

Introducing the BHPS and scalar five-quark models, 
we can now calculate numerically the intrinsic strange 
distribution in the proton using these models. 
Figure~\ref{pumpli} (top) shows the strange 
distribution from the BHPS model [Eq.~\eqref{1}] using 
$ m_p=0.938 $, $ m_1=m_2=m_3=m_p/3 $, and $ m_4=m_5=0.5 $ 
in GeV and also the obtained results from the scalar five-quark model 
[Eq.~\eqref{4}] using the exponential form factor [Eq.~\eqref{7}] with 
two values for the parameter $ \Lambda $ (2 and 5) and the same values for 
the physical masses. The strange distribution from the scalar five-quark 
model and using the power-law form factor with $ n=4 $ and the mentioned 
values for the physical masses and parameter $ \Lambda $ have been shown 
in Fig.~\ref{pumpli} (bottom) and compared with the BHPS result again. 
As can be seen from the figures, the obtained results from the BHPS and 
scalar five-quark model with the exponential form factor and $ \Lambda=2 $ 
are very similar, but the result related to the $ \Lambda=5 $ tended to the 
lower $ x $ and also is smaller and somewhat greater than the BHPS in the 
regions $ x<0.42 $ and $ x>0.42 $, respectively. This latter behavior is also 
seen when we use the scalar five-quark model with the power-law form factor 
[Eq.~\eqref{8}] and $ \Lambda=2 $ and $ 5 $. It should be noted that we 
have neglected the probability of the $\vert uud s\bar{s}\rangle $ state in the 
proton presently and normalized all curves so that the quark number 
condition $ \int_0^1 dxf(x)=1, (f=s, \bar s) $ is satisfied.
%
\section{Meson-Baryon Model}\label{MBM} 
As mentioned in the previous section, in the light-cone Fock 
space picture, the nucleon's wave function can be considered 
as a superposition of configurations of off-shell physical particles. 
So one of the phenomenological models for producing intrinsic 
quark distributions is the meson-cloud or meson-baryon
model. Unlike the five-quark models that lead to an equal 
distribution for strange and antistrange, this more dynamical 
model can lead to the $ s-\bar s $ asymmetry in the nucleon sea. 
Although the original meson-baryon model is rather complicated 
computationally \cite{Hobbs:2013bia, Melnitchouk:1992yd, 
Szczurek:1996ur, Kumano:1997cy, Traini:2013zqa}, we 
can consider a simple configuration as has been used in Ref.~\cite{Pumplin:2005yf}. 
According to the MBM, the nucleon can 
fluctuate to the virtual meson-baryon Fock states ($ N\longrightarrow MB $). 
For example, in the case of intrinsic strange we can consider the two-body 
state $ K^+\Lambda^0 $, where $ K^+ $ is a $ u\bar s $ meson and 
$ \Lambda^0 $ is a $ uds $ baryon. To calculate the intrinsic $ s $ and 
$ \bar s $ distributions in the nucleon, we should model the $ K^+\Lambda^0 $ 
probability distribution, $ uds $ distribution in $ \Lambda^0 $, and $ u\bar s $ 
distribution in  $ K^+ $ and then use the following relation defined as 
convolutions of the distributions \cite{Pumplin:2005yf}:
\begin{align}\label{9}
{{dP} \over {dx}} &= 
\int_0^1 \! dx_1 \, f_1(x_1) \int_0^1 \! dx_2 \, f_2(x_2) \, 
\delta(x - x_1 x_2)\nonumber\\ &=
\int_x^1{{dy} \over {y}} \, f_1(y) \, f_2(x/y).
\end{align}
   \begin{figure}[t!]
    \includegraphics[width=0.45\textwidth,clip]{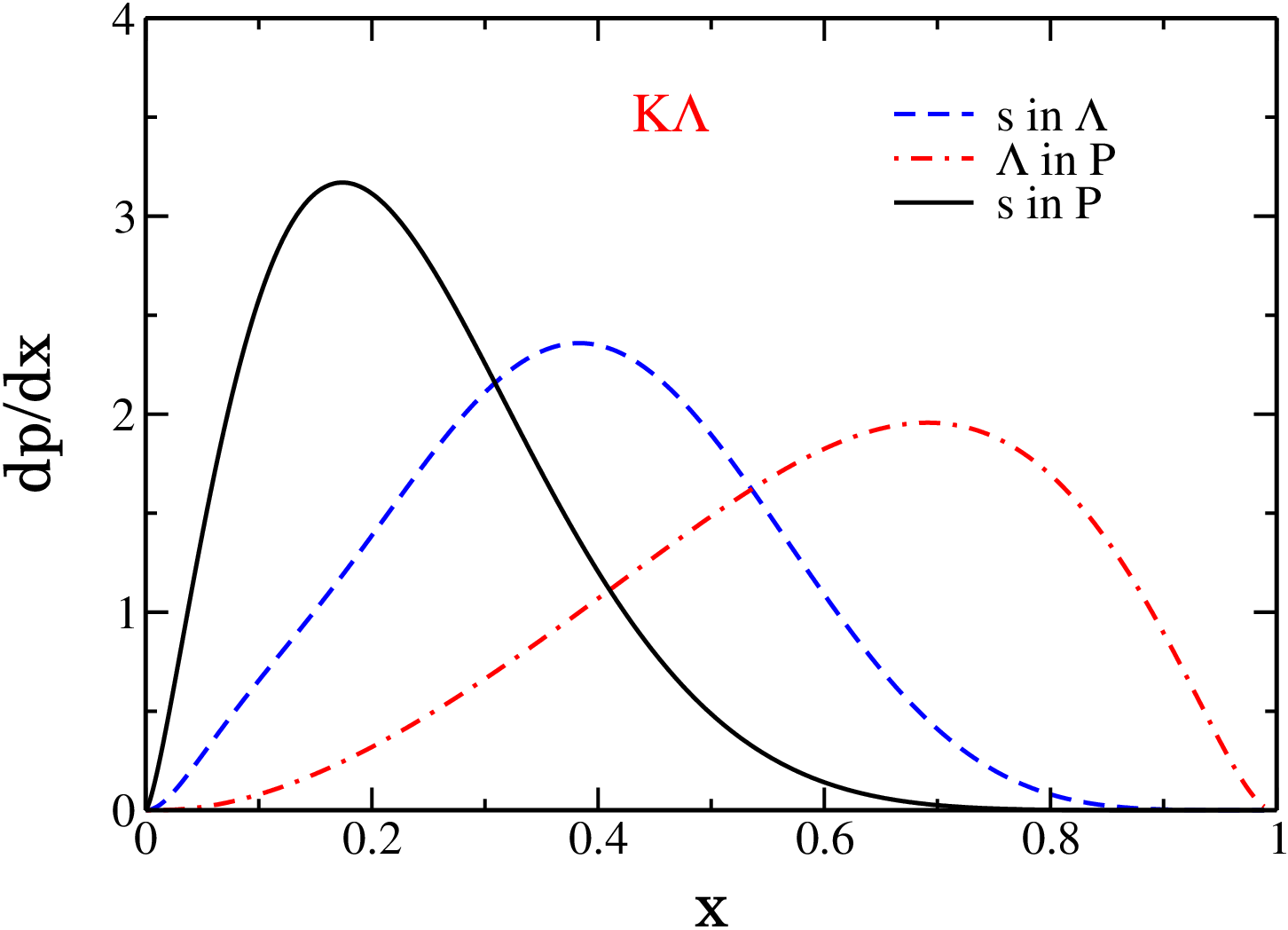}
    \includegraphics[width=0.45\textwidth,clip]{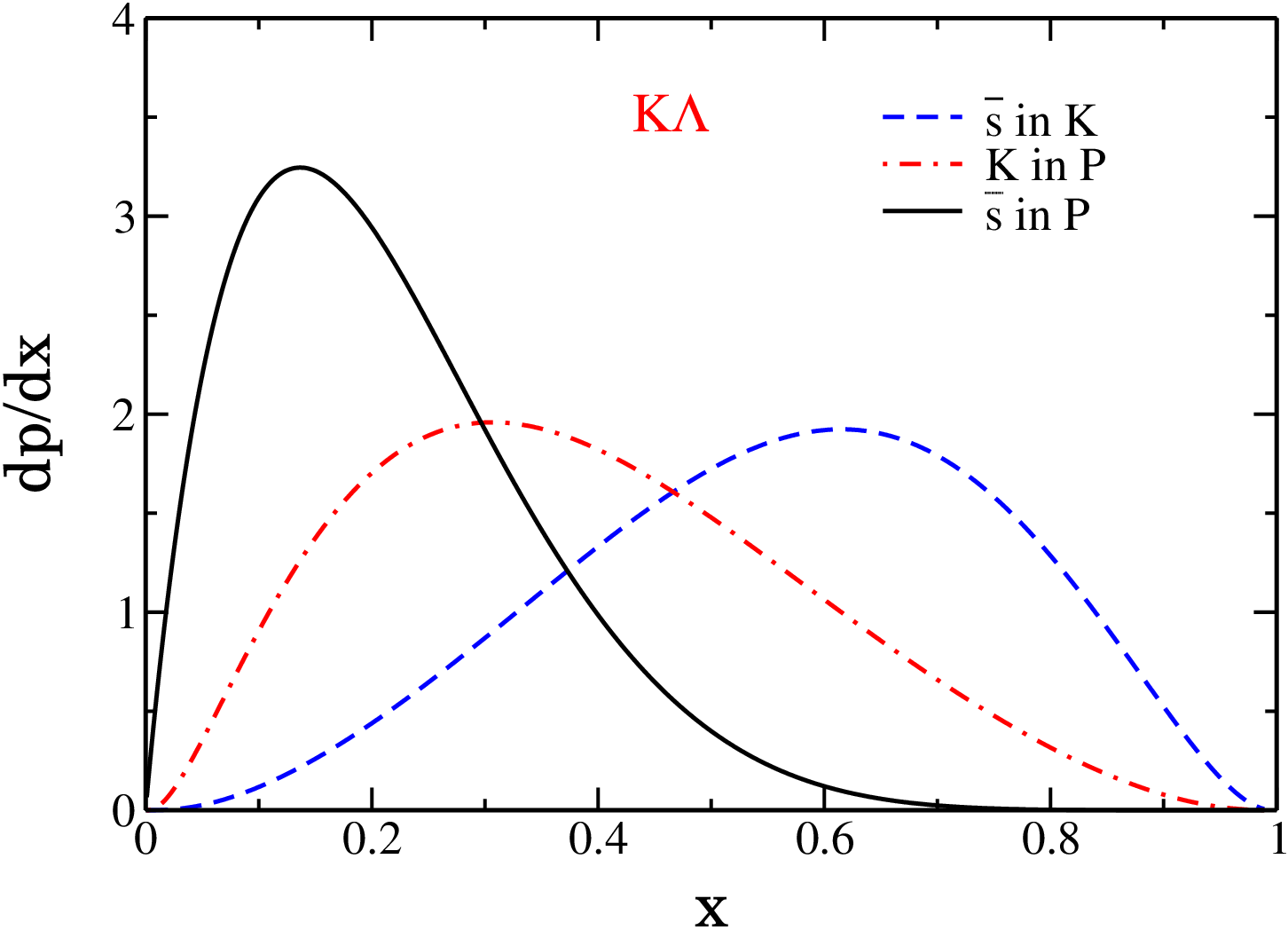}
    \caption{. Momentum distribution of the strange quark 
    in the $ \Lambda^0 $ baryon, $ \Lambda^0 $ from $ p\longrightarrow K^+
    \Lambda^0 $, and final $ s $ distribution in the proton (top). 
    Momentum distribution of the antistrange quark in the $ K^+ $ 
    meson, $ K^+ $ from $ p\longrightarrow K^+\Lambda^0 $, 
    and final $ \bar s $ distribution in the proton (bottom).}\label{mbmk}
\end{figure}

In this regard, we use Eq.~\eqref{4} with $ N=2 $ and 
$ F^2\varpropto (s_{K\Lambda}+\Lambda^2_{p})^{-2}$  
and $ F^2\varpropto (s_{u\bar s}+\Lambda^2_{K})^{-2}$ to 
model the  $ K^+\Lambda^0 $ probability distribution and $ u\bar s $ 
distribution in  $ K^+ $, respectively. To model the $ uds $ distribution in 
$ \Lambda^0 $ we have $ N=3 $ and $ F^2\varpropto 
(s_{uds}+\Lambda^2_{\Lambda})^{-2}$. Moreover, the required 
physical masses are chosen as $ m_K=0.4937 $, $ m_\Lambda=1.1157 $, 
$ m_p=0.938 $, $ m_s=m_{\bar s}=0.5 $, and $ m_u=m_d=m_p/3 $ in 
GeV. Figure 2 shows the obtained results for the $ s $ and $ \bar s $ 
distributions from $ p\longrightarrow K^+\Lambda^0 $ 
using $ \Lambda_p=4 $ and $ \Lambda_K=\Lambda_\Lambda=2 $. 
As can be seen, the $ \bar s $ distribution in the $ K^+ $ meson is harder 
than the $ s $ distribution in the $ \Lambda^0 $ baryon. Actually, since the 
fraction of the hadron mass carried by the constituent quark determines 
the position of the peak of quark distribution in $ x $, this difference is 
because the $ \bar s $ carries a larger fraction of the $ K^+ $ mass than the 
fraction of the $ \Lambda^0 $ mass carried by $ s $. Maybe we expect 
the same behavior for the strange and antistrange distributions in the proton. 
But in this case the momentum distribution of the meson or baryon in the 
two-body state $ MB $ also has an important rule. We see that by doing 
the convolution of Eq.~\eqref{9}, the $ s $ distribution in the proton is 
somewhat harder than the $ \bar s $ distribution.
   \begin{figure}[t!]
    \includegraphics[width=0.45\textwidth,clip]{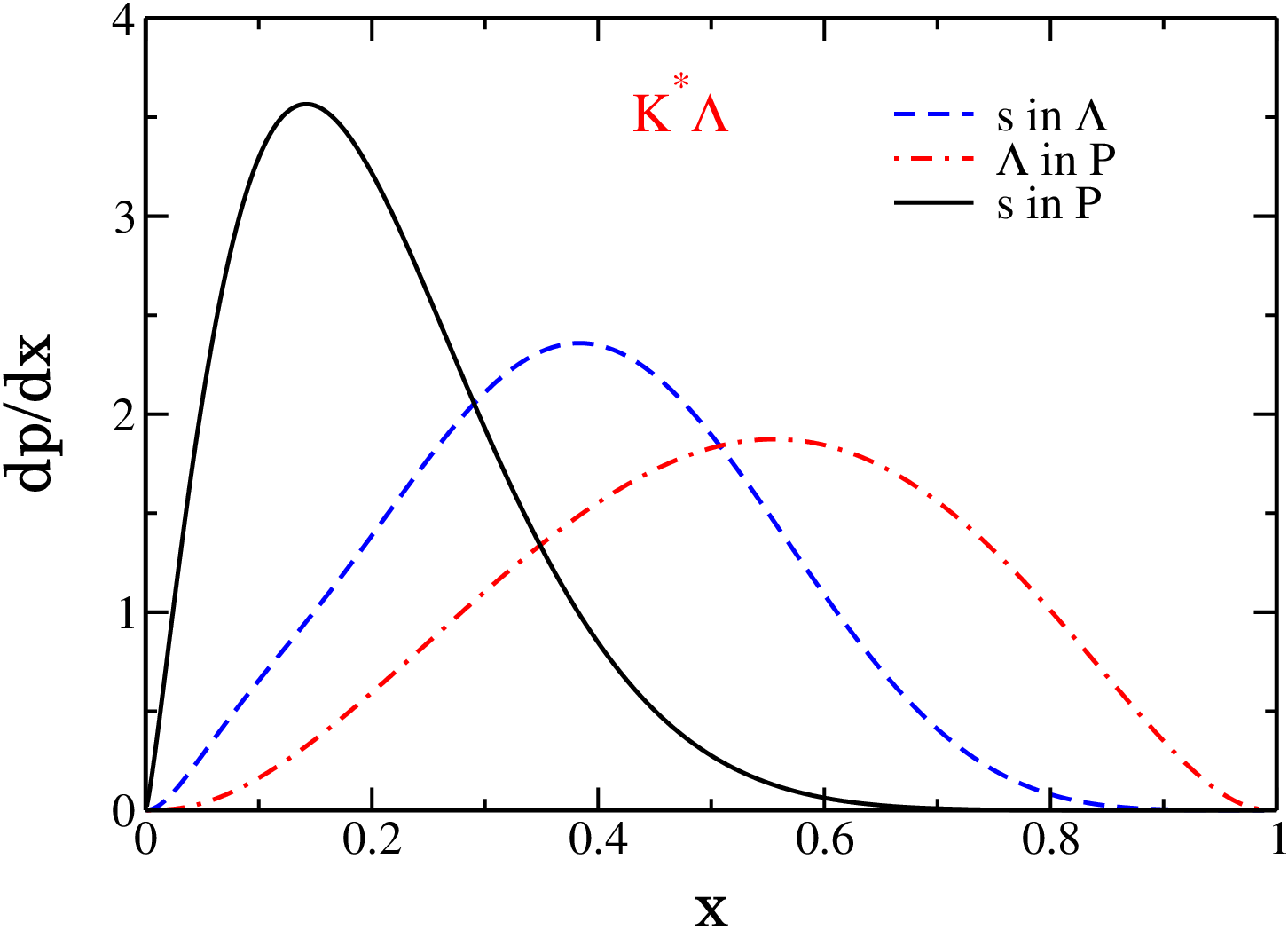}
    \includegraphics[width=0.45\textwidth,clip]{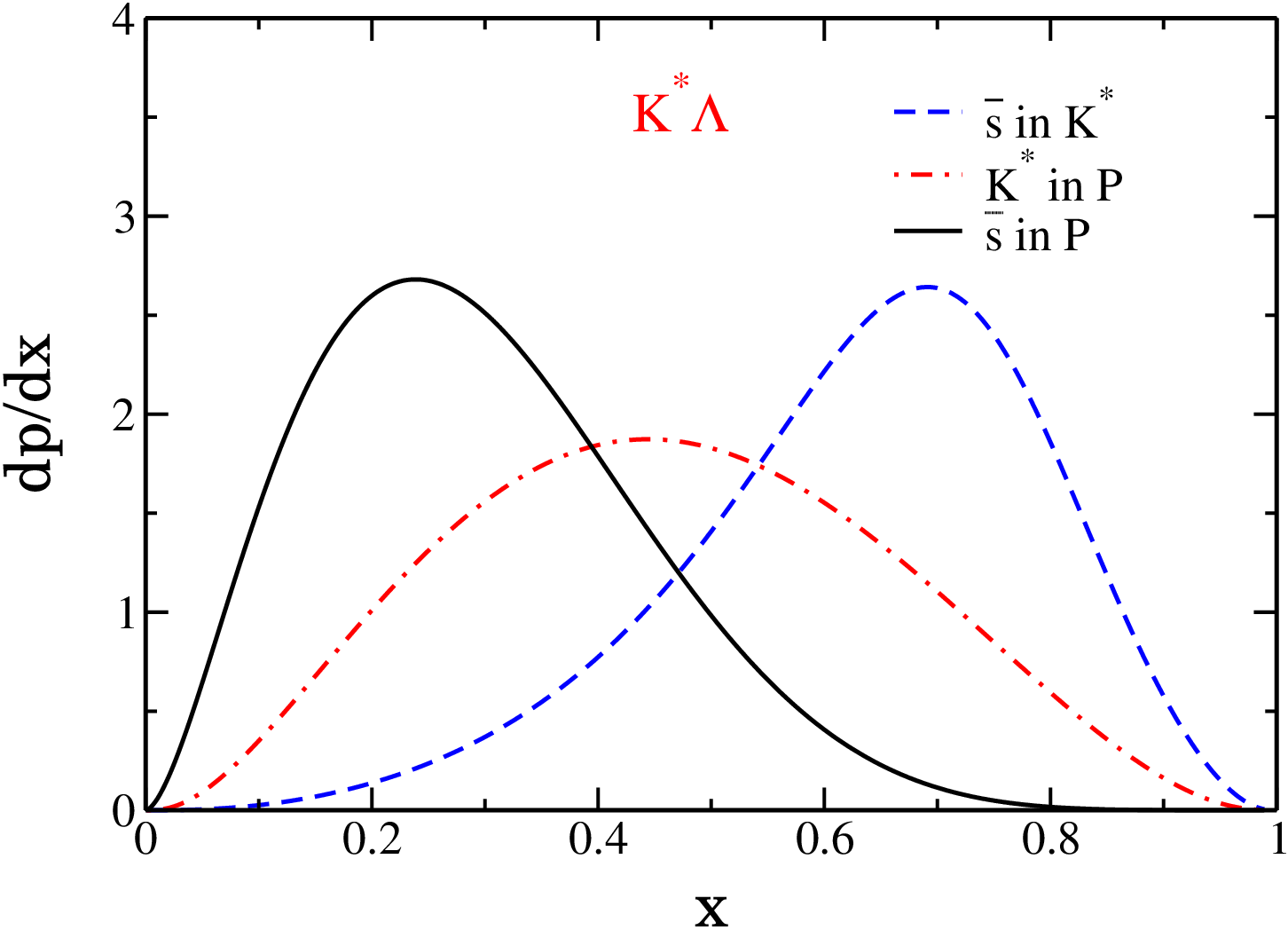}
    \caption{ Momentum distribution of the strange quark 
    in the $ \Lambda^0 $ baryon, $ \Lambda^0 $ from $ p\longrightarrow 
    K^{*+}\Lambda^0 $, and final $ s $ distribution in the proton (top). 
    Momentum distribution of the antistrange quark in the $ K^{*+} $ meson, 
    $ K^{*+} $ from $ p\longrightarrow K^{*+}\Lambda^0 $, and final 
    $ \bar s $ distribution in the proton (bottom).}
    \label{sbarink}
\end{figure}

In addition to the  $ K^+\Lambda^0 $, we can also consider the following fluctuations:
\begin{eqnarray}
p&\longrightarrow & K^0(d\bar{s})\Sigma^+ (uus),\nonumber\\
p&\longrightarrow & K^+(u\bar{s})\Sigma^0 (uds),\nonumber\\
p&\longrightarrow & K^{*+}(u\bar{s})\Lambda^0 (uds),\nonumber\\
p&\longrightarrow & K^{*0}(d\bar{s})\Sigma^+ (uus),\nonumber\\
p&\longrightarrow & K^{*+}(u\bar{s})\Sigma^0 (uds).
\label{10}
\end{eqnarray}
But since the physical masses of $ K^0 $ and $ K^+ $, $ K^{*0} $ 
and $ K^{*+} $, and even $ \Lambda $ and $ \Sigma $ are almost 
equal, the obtained results from some states are the same. Therefore we 
can consider only two states, $ K^+\Lambda^0 $ and $ K^{*+}\Lambda^0 $, 
that lead to the different shapes for $ s $ and $ \bar s $ distributions and thus 
the $ s-\bar s $ asymmetry. We can also sum the results obtained from these 
states. To calculate $ s $ and $ \bar s $ distributions from 
$ p\longrightarrow  K^{*+}\Lambda^0 $, we use the same physical 
masses as in the case $ K^+\Lambda^0 $, but for the antistrange 
in $ K^{*+} $ we take an effective mass $ m_{\bar s}=0.7 $ to keep 
$ m_{\bar s}+m_u>m_{K^*} $ and avoid mass singularity. Moreover, 
the $ K^{*+} $ mass is taken to be $ m_{K^*}=0.8917 $ GeV. The 
results for $ p\longrightarrow  K^{*+}\Lambda^0 $ are shown in 
Fig.~\ref{sbarink} using $\Lambda_p=4 $ and $ \Lambda_{K^*}=
\Lambda_\Lambda=2 $ again. Since we chose a larger mass for the
antistrange quark in the $ K^* $ meson than $ K $, the resulting $ \bar s $ 
distribution is even harder than before. One can see from Fig.~\ref{sbarink} 
that the momentum distribution of $ \Lambda^0 $ and $ K^{*+} $ from 
$ p\longrightarrow  K^{*+}\Lambda^0 $ are peaked almost around $ x=0.5 $, 
meaning that the meson  and baryon approximately share proton momentum 
fairly equally. So, because the antistrange quark carries a larger fraction of the
total momentum of the meson than the strange quark of the total momentum 
of the baryon, $ \bar s $ distribution in the proton is harder and has
a greater magnitude than the $ s $ distribution at large $ x $.

\begin{figure}[t!]
\centering
    \includegraphics[width=0.46\textwidth,clip]{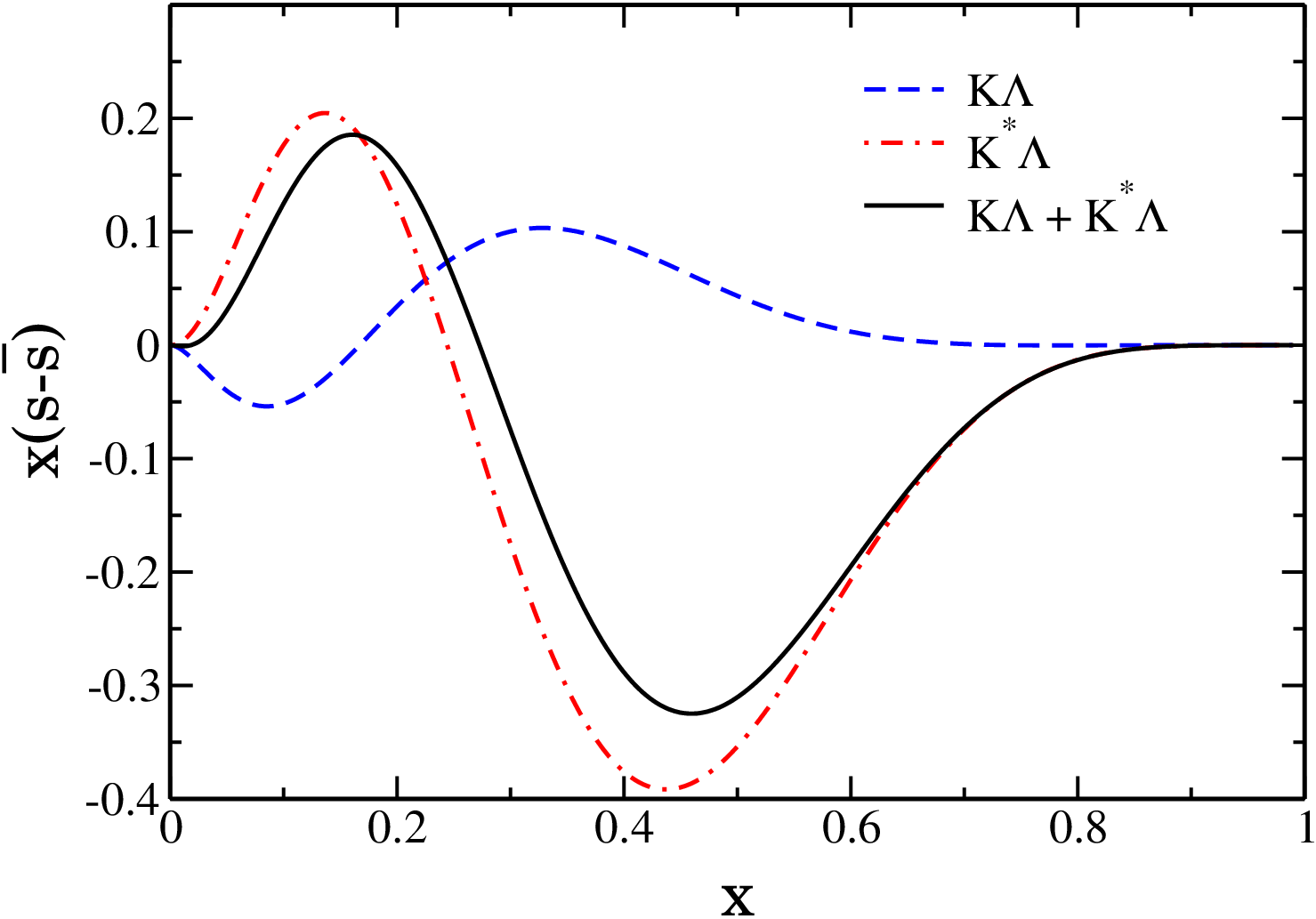}
    \caption{ The strange sea asymmetry $ x[s(x)-\bar s(x)] $ 
    calculated using the meson-baryon model from the fluctuation of the 
    proton to $ K\Lambda $ (dashed curve) and $ K^*\Lambda $ (dotted 
    dashed curve) states and also their summation (solid curve).}
    \label{32}
\end{figure}

As mentioned above, a main difference between the MBM and five-quark 
models is that the MBM can lead to the $ s-\bar s $ asymmetry in the 
nucleon sea. Actually, since in the MBM framework the probability 
distributions of the meson and baryon in the proton are different and 
$ s $ and $ \bar s $ also have different distributions in the baryon 
and meson respectively, the $ s-\bar s $ asymmetry is natural. 
The possibility of this asymmetry in the nucleon was discussed by 
Signal and Thomas \cite{Signal:1987gz} by applying MCM for the 
first time, and after that it has been investigated by other authors 
\cite{Brodsky:1996hc, Christiansen:1998dz, Cao:2003ny, 
Burkardt:1991di, Melnitchouk:1999mv}. Now, having $ s $ and 
$ \bar s $ distributions from $ p\longrightarrow K^+\Lambda^0 $ 
and $ p\longrightarrow  K^{*+}\Lambda^0 $, we can calculate strange 
and antistrange asymmetry for each of these states and also their sum in 
the proton. The results are shown in Fig.~\ref{32} and labeled 
as $ K\Lambda $, $ K^*\Lambda $, and $ K\Lambda+K^*\Lambda $, 
respectively. As can be seen, the results from $ K^+\Lambda^0 $ 
and $ K^{*+}\Lambda^0 $ are quite different. The summation of
 these two states is comparable with full MBM calculations \cite{Cao:2003ny}.

\section{The probability of intrinsic strange in the nucleon}\label{result} 

As mentioned in the introduction, several studies have been done so 
far to estimate the probability of the intrinsic charm in the nucleon (or 
momentum fraction carried by intrinsic charm) and have suggested different 
values for it. For example, BHPS estimate 1 \% probability for intrinsic 
charm in the proton from the diffractive production of charmed hadrons at 
large longitudinal momentum \cite{1981}. However, few analyses 
have been done to extract the probabilities of the light intrinsic quarks 
\cite{Chang:2011vx, Chang:2011du}. In order to calculate the probability 
of the $\vert uud s\bar{s}\rangle $, Chang and Pang used the existing 
$x(s+\bar{s})$ data from the HERMES Collaboration measurement of 
charged kaon production in a SIDIS reaction \cite{Airapetian:2008qf}. 
They suggested that these data contain both the extrinsic and intrinsic 
components of the strange sea that are dominant at small and large $ x $, 
respectively, and also found that data in the $x>0.1$  region can be described 
well by the intrinsic strange distributions from the BHPS model (see Sec.~\ref{brodsky}). 
They evolved the resulting distributions from the initial scale $\mu = 0.3$ GeV 
and $\mu = 0.5$ GeV to $Q^2=2.5$ GeV$^2$ and then extracted the normalization 
by fitting data with $x>0.1$, considering the assumption that the extrinsic sea 
component is negligible in this region. The obtained results for the probability 
of the $\vert uud s\bar{s}\rangle $ state are as follows \cite{Chang:2011du}:
\begin{align}\label{11}
\mathcal{P}^{s\bar{s}}_5&= 0.024 \qquad (\mu = 0.3\ \text{GeV}),\nonumber\\
\mathcal{P}^{s\bar{s}}_5& = 0.029\qquad (\mu = 0.5\ \text{GeV}).
 \end{align}
 
\begin{figure}[t!]
\centering
    \includegraphics[width=0.46\textwidth,clip]{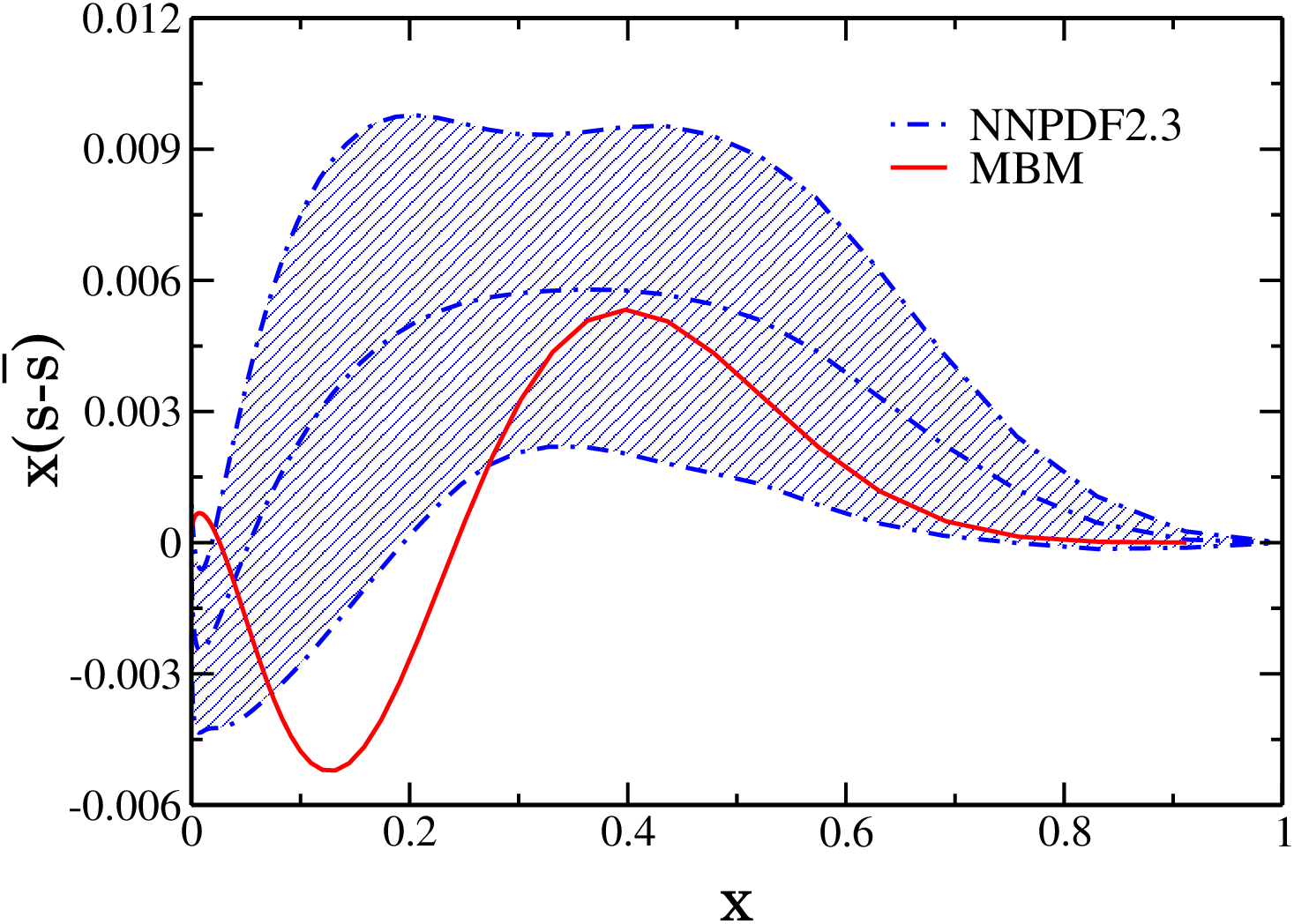}
    \caption{ The strange sea asymmetry $ x[s(x)-\bar s(x)] $ 
    calculated using the meson-baryon model from the fluctuation of the 
    proton to $ K\Lambda $ plus $ K^*\Lambda $ states and evolved from 
    $ \mu = 0.5 $ GeV to $ Q^2=16 $ GeV$ ^2$ (solid line) in comparison 
    with the NNPDF2.3 global fits \cite{Ball:2012cx} shown by the shaded areas.}
    \label{copmnnpdf}
\end{figure}

 As mentioned before, all distributions obtained in the previous sections 
 from the BHPS, scalar five-quark, and meson-baryon models have been 
 normalized to 100 \% probability so that the quark number 
 condition $ \int_0^1 dxf(x)=1, (f=s, \bar s) $ is satisfied. At this 
 stage, we can take the above-mentioned probability determined 
 using $ \mu = 0.5 $ and then evolve $ s $ and $ \bar s $ distributions 
 from the MBM (see the previous section) to calculate the $ x(s-\bar s) $ 
 asymmetry at $ Q^2=16 $ GeV$ ^2 $, for instance. The evolution of the 
 distributions can be carried out with the QCDNUM package \cite{Botje:2010ay}. 
 The result is shown in Fig.~\ref{copmnnpdf}  and also compared with 
 the result of NNPDF2.3 \cite{Ball:2012cx} for this quantity. 
 As can be seen, the MBM result for $ x(s-\bar s) $ is inside the 
 error bound of the NNPDF2.3 result in some regions of $ x $.
 It highlights this idea in our mind that using the purely theoretical 
 result of the MBM for the $ x(s-\bar s) $ asymmetry can be replaced 
 with a parametrization for this quantity in the global analysis of PDFs.
 
\section{Conclusion}\label{summery}
Since Brodsky \textit{et al.} suggested the possible existence 
of an intrinsic quark component in the nucleon, the intrinsic quark 
has been a subject of interest in many studies in hadron physics. 
These studies can be divided generally into two main categories: 
heavy and light intrinsic quarks. Although the heavy 
quarks have an important role in the study of many processes 
in the standard model and beyond it, one of the significant 
aspects of the nucleon structure is the distribution of strange and antistrange 
sea quarks and the possible asymmetry between them. More precise 
knowledge in this field is very important for better understanding of the 
nucleon structure and properties of the sea quarks and also for describing 
processes such as $ W $ boson production in association with 
charm jets \cite{Abazov:2014fka} or single top quark production 
\cite{He:2011ss}, as well as neutrino interactions \cite{Alberico:2001sd, Dore:2011qe}. 
Moreover, the $ s-\bar s $ asymmetry 
in the nucleon can also be very important for explaining experimental 
results such as the NuTeV anomaly reported by the NuTeV Collaboration 
\cite{Zeller:2001hh}. In this work, we concentrated on the intrinsic strange 
quark and calculated its distribution in the nucleon using various light-cone 
models. Actually, in addition to using the BHPS model \cite{Brodsky:1980pb, 
1981}, we used the scalar five-quark model and a simple meson-baryon 
model introduced by Pumplin \cite{Pumplin:2005yf} to calculate the intrinsic 
strange distribution in the nucleon numerically and then compared the results 
with each other. We found that these models can leads to the rather different 
distributions for the intrinsic strange that can be dominated in different 
values of $ x $. The resulting distributions from the scalar five-quark model in 
some situations are very similar to the BHPS result, but in some cases tended to 
the lower $ x $ and also are somewhat greater than the BHPS in larger $ x $. 
Although the BHPS and scalar five-quark models cannot give us any asymmetry 
between the $ s $ and $ \bar s $ distributions in the nucleon, the MBM 
leads to the $ s-\bar s $ asymmetry. This is a very important conclusion because 
the perturbative (extrinsic) sea quark distributions in the nucleon are clearly 
symmetric, so the flavor asymmetry in the nucleon certainly has a 
nonperturbative origin. We compared the obtained result for this asymmetry 
to the NNPDF2.3 \cite{Ball:2012cx} result, considering the extracted 
probability in Ref. \cite{Chang:2011du} for the intrinsic strange in the nucleon, 
and found that they are in good agreement with each other. Therefore, maybe 
using the purely theoretical result of the MBM for 
the $ x(s-\bar s) $ asymmetry can be replaced with a parametrization for this 
quantity in the global analysis of PDFs.
\acknowledgments
The author is grateful to H. Khanpour for reading the manuscript 
and useful discussions and comments. This project was
financially supported by Islamic Azad University-Kerman Branch.
\appendix
\section*{Appendix: Parametrization form for the intrinsic strange}
We presented various light-cone models for the intrinsic strange in 
the nucleon and discussed them in detail. As a next work 
related to the intrinsic strange quark, we provide a simple parametric 
form for all strange distributions in the nucleon computed in the light-cone 
framework using the BHPS, scalar five-quark, and MBM models. 
The parametric form is taken to be
 \begin{align}\label{12}
f(x)&=Ax^b(1-x)^c.
\end{align}
The normalization constant $A=1/B(1+b,1+c)$, where $B$ is the Euler 
beta function, obtained from the quark number sum rule $\int_0^1 f(x) dx=1$ 
and ensure that the distributions are normalized to $1$. The best-fit parameter 
values are listed in Table~\ref{para}.
\begin{table}[b!]
\caption{Best-fit parameter values for the intrinsic strange.}
\centering
\begin{tabular}{cccc}			\hline\hline
Model&$a$&$b$&$c$\\ \hline
BHPS&\qquad 2.265$\times10^2$&\qquad 8.433&\qquad 1.449\\
Exponential ($\Lambda = 2$)&\qquad 3.348$\times10^2$&\qquad 9.015&\qquad 1.592\\
Power law ($\Lambda = 2$)&\qquad 8.514$\times10^1$&\qquad 7.297&\qquad 1.053\\
Exponential ($\Lambda = 5$)&\qquad 3.355$\times10^1$&\qquad 6.124&\qquad 0.700\\
Power law ($\Lambda = 5$)&\qquad 3.028$\times10^1$&\qquad 5.905&\qquad 0.673\\
MBM ($s$ from $K\Lambda $)&\qquad 1.350$\times10^2$&\qquad 6.704&\qquad 1.413\\
MBM ($\bar{s}$ from $K\Lambda $)&\qquad 5.549$\times10^1$&\qquad 6.145&\qquad 0.973\\
MBM ($s$ from $K^*\Lambda $)&\qquad 1.404$\times10^2$&\qquad 7.713&\qquad 1.277\\
MBM ($\bar{s}$ from $K^*\Lambda$)&\qquad 1.246$\times10^2$&\qquad 5.317&\qquad 1.667\\
\hline\hline
\end{tabular}
\label{para}
\end{table}

\end{document}